\documentclass[aps,preprint,showpacs]{revtex4}
\begin{document}        %
\draft
\title{Note on "Sonic Mach cones induced by fast partons in a perturbative quark-gluon plasma"}
\author{Zotin K.-H. Chu}  
\affiliation{3/F, 24, 260th. Lane, First Section, Muzha Road,
Wenshan District, Taipei, Taiwan 116, China}
\begin{abstract}
We make remarks on  Neufeld {\it et al.}'s [{\it Phys. Rev. C} 78,
041901(R) (2008)] paper especially about the Mach cone formation.
We argue that the original bow shock structure (as a fast parton
moving  through a quark-gluon plasma) has been smeared out after
the approximations made by Neufeld {\it et al.}
%
\end{abstract}
\maketitle
\bibliographystyle{plain}
Neufeld {\it et al.} recently presented a solution obtained from
the linearized hydrodynamical equations of the medium (a fast
parton traversing a weakly coupled quark-gluon plasma) in which it
contains a sonic Mach cone and a dissipative wake if the parton
moves at a supersonic speed [1]. In fact, Casalderrey-Solana {\it
et al.} [2] showed that if one couples an appropriately chosen
supersonic sound source to a linearized hydrodynamical evolution,
one can obtain a propagating Mach cone (Casalderrey-Solana has
stressed that the conical flow scenario for the observed shape is
a consequence of the emission of sound by a supersonic high
momentum particle propagating in the quark gluon plasma [3]). The
present author, however, based on some previous works [4-5], likes
to argue here that the Mach cone interpretation in [1] is not
complete.
\newline
In fact, the present author argues that there are shock waves
occurring under the conditions discussed in [1] as the medium
through which the parton propagates  is not dilute but dense
[2,6]. As noted in [2], the moving 'undressed' hard parton being
constantly emitting gluons, which emit new ones etc., and thus the
whole shower (or the core) is a complicated nonlinear phenomenon
(the multiplicity of this shower grows nonlinearly with time) and
the combination should obviously be treated   as a macroscopic
body [7-8] passing through the medium. This could be traced in
[2]:
{\it Since the velocity of the shock depends on its intensity, the
cone should in fact be somewhat rounded near its top. This effect
is ignored $\cdots$} (cf. the note or Ref. 9 in [2]) or {\it The
region near the head of the jet, which we will refer to as a
'non-hydrodynamical core', $\cdots$ As found in [4]}(R. Baier,
Y.L. Dokshitzer, S. Peigne  and D. Schiff, Phys. Lett. B {\bf
345}, 277 (1995);  R. Baier, Y.L. Dokshitzer, A.H. Mueller and D.
Schiff, JHEP {\bf 20010109} 033 (2001)), {\it the multiplicity of
this shower grows nonlinearly with time, so eventually the core
may become a macroscopic body, providing a large perturbation of
the matter. From the hydrodynamical point of view, its size is
limited from below by the dissipative 'sound attenuation length'
$\Gamma_s = (4/3)\eta/(\epsilon + p)$, with $\eta$ being the shear
viscosity} (page 24 of [2]; please see the details for the
relevant symbols therein).\newline
Meanwhile, there is inconsistency in the theoretical treatment in
[1] which will be described below. Neufeld {\it et al.} solved the
hydrodynamical equation (cf. Eq. (8) and the detailed explanations
for each symbol in [1])
\begin{equation}
 \partial_{\mu} T^{\mu\nu}= J^{\nu} \equiv \int \frac{d {\bf
 p}\,p^{\nu}}{(2\pi)^3} (\nabla_{p_i} D_{ij}({\bf p},t) \nabla_{p_j}
 f({\bf x},{\bf p},t))
\end{equation}
by assuming that the energy and momentum density deposited by the
{\bf parton} is small compared to the equilibrium energy density
of the medium. Here, $f({\bf x},{\bf p}, t)$ is the ensemble
averaged phase-space distribution of medium partons [1] and
\begin{equation}
   D_{ij} ({\bf p},t)=\int_{-\infty}^t dt' F_i ({\bf x},t)F_j
   ({\bf x'},t'),
\end{equation}
with $F_i({\bf x},t)=g Q^a(t)[E_i^a({\bf x},t)+({\bf v}\times {\bf
B})^a_i ({\bf x},t)]$ being the color Lorentz force on a medium
particle and
 $J^{\nu}$ represents a source term due to the interaction of
the medium with the passing fast parton. Subsequently, to evaluate
the source term in the right-hand-side of Eq. (1), Neufeld {\it et
al.} considered a thermal plasma of massless gluons with the
unperturbed distribution (cf. Eq. (12) in [1]). The present author
doubts : Where is the contribution from the {\bf entire parton}
(say, quarks) [2,9-11]?
\newline
To briefly check the mathematical derivations of [1] or [10] (they
were self-cited, cf. Ref. 24 in [10], i.e.,  [1] here or Ref. 19
in [1], i.e., [10] here), we start from equations (13) : Integral
form of $J^0 ({\bf x},t)$ and (14) : Integral form of $J^k ({\bf
x},t)$. We have no idea how the term : $i \epsilon$ can be
inserted into the denominator of the integrand of above mentioned
integrals (cf. $(\omega'-{\bf k}'\cdot \hat{{\bf v}}+i \epsilon)$
in Eqs. (13) and (14) of [1])? As the first author for [1] and
[10] is the same, then the present author tried to trace this back
from [10]. Similarly, we also have no idea how the term : $i
\epsilon$ can be inserted into the denominator of the integrand of
the integral form for (i.e., Eq. (12) in [10])
\begin{equation}
  f_1^a=-\frac{i g C_2}{N_c^2-1}\int \frac{d^4k}{(2\pi)^4}
  \int d^4 x' U_{ab}(x,x')\times \frac{e^{ik(x'-x)}}{v\,k+i\epsilon}
  {\bf F}^b (x')\cdot \nabla_p f_0,
\end{equation}
\begin{displaymath}
 f_1^a \equiv f^a ({\bf x},{\bf p},t)=\int dQ Q^a f({\bf x},{\bf
 p},t).
\end{displaymath}
Please refer the detailed explanation for symbols or notations
appeared in above expression to [10]. However, there are no
definition for $i \epsilon$ up to the relevant statements near Eq.
(12) of [10]? Even though {\it A detailed derivation of the source
term, including color screening by the medium, is presented in
Ref. [19]} appeared in [1]?.
\newline
Meanwhile we know that Mach cones are V-shaped disturbances
produced by a supersonic object [7-8] or the interference of sound
waves from a supersonic source leads to the Mach cone [3]. They
are familiar in gas dynamics [7-8]. The cone's Mach angle is
$\theta= \sin^{-1} 1/M$, where $M=u/c$ is the Mach number of an
object moving at speed $u$ through a medium with an acoustic speed
$c$. As we argued above that there are shock waves appearing in
[1]. The disappearing of shock waves in [1]  could be traced in
the following : {\it Since hydrodynamics is only valid at
distances that are large compared to the mean free path, and in a
weakly coupled plasma the mean free path is parametrically large
compared to the color screening length, the source term generated
by an energetic parton is, in first approximation, point-like. In
this spirit, the source term [Eqs. (15)-(18)] derived here can be
thought of as a sophisticated representation of a $\delta$
function $\cdots$}. It means the strong nonlinearity due to the
propagating blunt-body-like [7] parton has been smeared out due to
point-like treatments in [1]. Similar smearing-out could be traced
by {\it ... We evaluate each term by boosting to a frame comoving
with our volume element and then exploiting the assumption of
local thermal equilibrium.} Note that near the shock as there is a
discontinuity and entropy condition [5,7], it is not at (thermal)
equilibrium. \newline The originally curved {\bf bow shocks}
[7-8,12] have been replaced and approximated by linear V-shaped
(Mach) wave patterns [2,7-8]. Furthermore, to remind the readers,
as mentioned in [1] : {\it We will incorporate the effect of color
screening and short-distance quantum effects by appropriate
infrared and ultraviolet cutoffs.}, considering above both
statements, how can the authors of [1] capture the detailed
long-wave (larger length-scale) limit or hydrodynamical behavior
(say, shock structures) by only a short-distance (smaller
length-scale) treatment and simultaneously neglecting the highly
nonlinearities (we well as the singularities due to the core or
'macroscopic body' [2]) by using a set of cutoffs?
\newline Finally, to be precise, considering the analogy from the
dusty plasma (cf. Samsonov {\it et al.} in [8]), due to the finite
size of the Debye sphere surrounding the fast-moving parton, the
vertex of the Mach cone is rounded rather than pointed. From
figures 1, 2 and 3 of [1], we cannot observe this Debye sphere?
The important information about the opening angle of the Mach cone
[8] is also absent in [1].

\end{document}